\documentclass[aps,prd,preprint,groupedaddress]{revtex4-1}

\usepackage{graphicx}
\usepackage{epsfig}
\usepackage{amsmath}
\usepackage{amsfonts,amssymb}
\usepackage{color}

  \usepackage{slashed}
  \usepackage{url}
\usepackage{bm}
\usepackage{verbatim}
\usepackage{float}

\usepackage{mathtools}

\usepackage{silence}
\begin{document}


\title{Differential Dyson-Schwinger equations for quantum chromodynamics}


\author{Marco Frasca}
\email[Marco Frasca\ ]{marcofrasca@mclink.it}
\affiliation{Via Erasmo Gattamelata, 3 \\
							00176 Rome (Italy)}


\date{\today}

\begin{abstract}
Using a technique devised by Bender, Milton and Savage, we derive the Dyson-Schwinger equations for quantum chromodynamics in differential form. We stop our analysis to the two-point functions. The 't~Hooft limit of color number going to infinity is derived showing how these equations can be cast into a treatable even if approximate form.
It is seen how this limit gives a sound description of the low-energy behavior of quantum chromodynamics by discussing the dynamical breaking of chiral symmetry
and confinement, providing a condition for the latter.
This approach exploits a background field technique in quantum field theory.
\end{abstract}


\maketitle


\section{Introduction}

The main difficulty of quantun chromodynamics (QCD) is that, at low energies, the theory is not amenable to treatment using perturbation techniques. This implies that some non-perturbative methods should be devised to solve them. The most widespread approach is solving the equations of the theory on a large lattice using computer facilities. This permitted to obtain, with a precision of a few percent \cite{Fodor:2012gf,pdg}, some relevant observables of the theory. This method improves as the computer resources improve making even more precise the comparison with experiment. Use of numerical techniques is a signal that we miss some sound theoretical approach to compute observables. 

A similar situation is seen for the correlation functions of the theory. Studies on the lattice of the gluon and ghost propagators, mostly in the Landau gauge, \cite{Bogolubsky:2007ud,Cucchieri:2007md,Oliveira:2007px} and the spectrum \cite{Lucini:2004my,Chen:2005mg} proved that a mass gap appears in a non-Abelian gauge theory without fermions. Theoretical support for these results was presented in \cite{Cornwall:1981zr,Cornwall:2010bk,Dudal:2008sp,Frasca:2007uz,Frasca:2009yp,Frasca:2015yva} providing closed form formulas for the gluon propagator. Quite recently, the set of Dyson-Schwinger equations for this case was solved, for the 1- and 2-point functions, and the spectrum very-well accurately computed both in 3 and 4 dimensions \cite{Frasca:2016sky,Frasca:2015yva,Frasca:2017slg}. Confinement was also proved to be a property of the theory \cite{Frasca:2016sky,Chaichian:2018cyv}.

Indeed, the Dyson-Schwinger equations were considered, since the start, the most sensible approach to treat a non-perturbative theory like QCD at low-energies \cite{Baker:1976vz,Eichten:1974et,Roberts:1994dr} and, more recently, \cite{Alkofer:2000wg}. In any case, the standard technique is to reduce the set of equations, that normally are partial differential equations, to their integral form in momentum space. Some years ago, Bender, Milton and Savage \cite{Bender:1999ek} proposed to derive the Dyson-Schwinger equations and treat them into differential form. This way to manage these equation was the one used to find the exact solution \cite{Frasca:2015yva}. This technique appears more general as it permits to work out a solution to a quantum field theory also when a background field is present. This is a rather general situation when a non-trivial solution of the 1-point equation is considered. Such a possibility opens up the opportunity of a complete solution to theories that normally are considered treatable only through perturbation methods. The idea is that, knowing all the correlation functions, a quantum field theory is completely solved.

The aim of this paper is to derive the Dyson-Schwinger equations for QCD in differential form. We obtain them for the 1- and 2-point functions. We show that, in the 't~Hooft limit \cite{tHooft:1973alw}, the equation can be cast into a treatable form. It appears that a non-local Nambu-Jona-Lasinio model is the proper low-energy limit of the theory \cite{Frasca:2016rsi}.

The paper is so structured. In Sec.~\ref{sec2}, we give the main equations and notations. In Sec.~\ref{sec2.0}, we presente the technique we will use to obtain the Dyson-Schwinger set of equations. In Sec.~\ref{sec2a}, we derive the set of equations for the 1- and 2-point correlation functions. In Sec.~\ref{sec3}, we discuss the 't~Hooft limit. In Sec.~\ref{sec4a}, we discuss the dynamical breaking of the chiral symmetry and derive a confinement condition. Finally, in Sec.~\ref{sec5}, we present the conclusions.

\section{Basic equations\label{sec2}}

In QCD, one has the Lagrangian (to fix the notation we assume $(1,-1,-1,-1)$ for the metric signature)
\begin{equation}
\label{eq:L1}
{\cal L}={\cal L}_{{inv}}+{\cal L}_{{gf}}+{\cal
L}_{{FP}}
\end{equation}
where ${\cal L}_{{inv}}$ denotes the classical gauge-invariant part, ${\cal L}_{{gf}}$ the gauge-fixing terms and ${\cal L}_{{FP}}$ the Faddeev-Popov (FP) ghost term characteristic of non-Abelian gauge theories \cite{Smilga:2001ck}
\begin{eqnarray}
\label{eq:L2} 
{\cal L}_{{inv}}&=&-\frac{1}{4}F_{\mu\nu}\cdot
F^{\mu\nu}-\bar\psi(i\gamma_\mu D^\mu+m)\psi\,,\cr
%
{\cal L}_{{gf}}&=&-\frac{1}{2\alpha}(\partial\cdot A)^2\,,\cr
{\cal L}_{{FP}}&=& -\bar c\cdot\partial_\mu D^\mu c\,.\label{lagr_terms} 
\end{eqnarray}
in the usual notation. We set $\alpha$ for the gauge parameter and $D_\mu$ represents the covariant derivative whose explicit forms are given by
\begin{eqnarray} 
D_\mu\ \psi&=&(\partial_\mu-igT\cdot A_\mu)\psi\,,\cr
D_\mu\ c&=&\partial_\mu c+gA_\mu\times c\,.\label{cov_deriv}
\end{eqnarray}

\section{Dyson-Schwinger equations \label{sec2.0}}

It is a rather old result that a tower of equations for the correlation functions of a quantum field theory can be obtained from the equations of motion \cite{Dyson:1949ha,Schwinger:1951ex,Schwinger:1951hq}. The solutions of this set of equations, when performed exactly, solves completely the theory itself. So, it appears like a very powerful non-perturbative approach. Anyway, the set of equations is such that any given equation depends on the correlation functions of higher order making a truncation needed. The set can be obtained using functional techniques as shown in \cite{Roberts:1994dr,Alkofer:2000wg} and, more recently, in \cite{Bender:1999ek}. We are going to describe both the methods to derive them with a simple example while this paper will rely on the latter. It is important to emphasize that, whatever approach one takes, the starting point are always the classical equations of motion satisfying a given variational principle.  

\subsection{Standard technique}

The general starting point is the partition function, for a theory with action $S[\phi]$,
\begin{equation}
Z[j]={\cal N}\int [d\phi]e^{iS[\phi]-i\int d^4xj(x)\phi(x)}
\end{equation}
being $\phi(x)$ a scalar field. The idea is that this functional integral does not change after a reparametrization $\phi(x)\rightarrow\phi(x)+\alpha(x)$, being $\alpha(x)$ an arbitrary function. Then, this gives
\begin{equation}
Z[j]\rightarrow Z[j]\langle e^{i\int d^4x\alpha(x)\left(\frac{\delta S}{\delta\phi(x)}-j(x)\phi(x)\right)}\rangle_j,
\end{equation}
from which, by requiring invariance, we derive the quantum equation of motion
\begin{equation}
\label{eq:dsder}
\left\langle\frac{\delta S}{\delta\phi(x)}\right\rangle_j=j(x).
\end{equation}
The procedure can be iterated taking further derivative with respect to $j$ and this will give all the set of Dyson-Schwinger equations for the correlation functions. We will obtain the final result by taking $j=0$ at the end of the computation. We note that from the lhs of eq.~(\ref{eq:dsder}) one gets the average on the classical equations of motion of the theory that are the starting point of the procedure.

From this point on, the standard technique introduce the vertex functions $\Gamma_n(\phi)$ using a Legendre transform. This means that, for a simple $\phi^4$ theory, we will get the first two Dyson-Schwinger equations
\begin{eqnarray}
\partial^2G_1&+&m^2 G_1+\lambda G_1^3+3\lambda G_2(0)+\lambda\int d^4zd^4z'd^4z''G_2(x-z)G_2(x-z')G_2(x-z'')\Gamma_3(z,z,',z'')=0 \nonumber \\
G_2^{-1}(x-y)&=&(\partial^2+m^2)^{-1}\delta^4(x-y)+3\lambda G_2(0)\delta^4(x-y) \nonumber \\
&+&\lambda\int dzdz'dz''G_2(x-z)G_2(x-z')G_2(x-z'')\Gamma_4(z,z,',z'',y).
\end{eqnarray}
Written in this way, the only viable approach is through a Fourier transform to momenta and working with integral equations. This could hide the physical content of the equations and the deep meaning of a possible truncation.

\subsection{Bender-Milton-Savage method}

Bender-Milton-Savage method \cite{Bender:1999ek} takes the move from eq.~(\ref{eq:dsder}), again the average of the classical equations of motion, but does not take any Legendre transform so, vertex functions are not introduced at all. Rather, one always works with higher-order n-point functions $G_n(x_1,x_2,\ldots,x_n)$. This permits to preserve the differential structure of the Dyson-Schwinger equations making them particular useful when exact solutions are known. This will give for a $\phi^4$ theory \cite{Frasca:2015yva}
\begin{eqnarray}
&&\partial^2G_1+m^2 G_1+\lambda G_1^3+3\lambda G_2(0)+\lambda G_3(0,0)=0 \nonumber \\
&&(\partial^2+m^2)G_2(x-y)+3\lambda G_2(0)G_2(x-y)+3\lambda[G_1(x-y)]^2G_2(x-y) \nonumber \\ 
&&+3\lambda G_3(0,x-y)G_1(x)+\lambda G_4(0,0,x-y))= \delta^4(x-y).
\end{eqnarray}
In this case, we have non-trivial exact solutions for $G_1$, being $G_3(0,0)=0$ and the the set of Dyson-Schwinger equations becomes treatable without any truncation. This would not have been possible with the Dyson-Schwinger equations written in the integral form as demanded by the standard technique.

\section{Correlation functions in QCD\label{sec2a}}

As already stated above, we will use the Bender-Milton-Savage method.
%
In order to get the Dyson-Schwinger equations, we have to start from the classical equations of motion, given by the functional derivatives of the action as in eq.~(\ref{eq:dsder}), and average on them. These are given by
\begin{eqnarray}
   &&\partial^\mu\partial_\mu A^a_\nu-\alpha^{-1}\partial_\nu(\partial\cdot A^a)+gf^{abc}A^{b\mu}(\partial_\mu A^c_\nu-\partial_\nu A^c_\mu)+gf^{abc}\partial^\mu(A^b_\mu A^c_\nu)+g^2f^{abc}f^{cde}A^{b\mu}A^d_\mu A^e_\nu \nonumber \\
	&&= gf^{abc}\partial_\nu(\bar c^b c^c) + g\sum_{q,i}{\bar q}^i\gamma_\nu T^a q^i+j_\nu^a \nonumber \\
	 &&\partial^\mu\partial_\mu c^a+gf^{abc}\partial^\mu(A_\mu^bc^c)=\varepsilon^a \nonumber \\
	&&(i\slashed\partial+g{\bm T}\cdot\slashed{\bm A}-m_q)q^i=\eta_q^i.
\end{eqnarray}
Letters $a,b,c,\ldots=1,\ldots, 8$ are for gluons and $i,j,k,l,\ldots =1,2,3$ are for colors. The scalar product runs on the former. Flavors are identified with the letter $q=u,d,s,c,t,b$. We fix the gauge to the Landau gauge, $\alpha\rightarrow 0$, and $c,\ \bar c$ are the ghost fields. Averaging on the vacuum state and dividing by the partition function $Z_{QCD}[j,\bar\varepsilon,\varepsilon,\bar\eta_q,\eta_q]$ one has
\begin{eqnarray}
    &&\partial^2G_{1\nu}^{(j)a}(x)+
		gf^{abc}(\langle A^{b\mu}\partial_\mu A^c_\nu\rangle-\langle A^{b\mu}\partial_\nu A^c_\mu\rangle)Z^{-1}_{QCD}[j,\bar\varepsilon,\varepsilon,\bar\eta_q,\eta_q]
		+gf^{abc}\partial^\mu\langle A^b_\mu A^c_\nu\rangle Z^{-1}_{QCD}[j,\bar\varepsilon,\varepsilon,\bar\eta_q,\eta_q]
		\nonumber \\
		&&+g^2f^{abc}f^{cde}\langle A^{b\mu}A^d_\mu A^e_\nu\rangle Z^{-1}_{QCD}[j,\bar\varepsilon,\varepsilon,\bar\eta_q,\eta_q] 
		=gf^{abc}\langle\partial_\nu(\bar c^b c^c)\rangle Z^{-1}_{QCD}[j,\bar\varepsilon,\varepsilon,\bar\eta_q,\eta_q] \nonumber \\
		&&+g\sum_{q,i}\langle{\bar q}^i\gamma_\nu T^a q^i\rangle  Z^{-1}_{QCD}[j,\bar\varepsilon,\varepsilon,\bar\eta_q,\eta_q]+ j_\nu^a \nonumber \\
	 &&\partial^2 P^{(\varepsilon)a}_1(x)
	 +gf^{abc}\partial^\mu\langle A_\mu^bc^c\rangle Z^{-1}_{QCD}[j,\bar\varepsilon,\varepsilon,\bar\eta_q,\eta_q]=\varepsilon^a \nonumber \\
	 &&(i\slashed\partial-m_q)q_{1}^{(\eta)i}+g{\bm T}\cdot\langle\slashed{\bm A}q^i\rangle Z^{-1}_{QCD}[j,\bar\varepsilon,\varepsilon,\bar\eta_q,\eta_q]=\eta_q^i.
\end{eqnarray}
The one-point functions are given by
\begin{eqnarray}
    &&G_{1\nu}^{(j)a}(x)Z_{QCD}[j,\bar\varepsilon,\varepsilon,\bar\eta_q,\eta_q]=\langle A^a_\nu(x)\rangle \nonumber \\
		&&P^{(\varepsilon)a}_1(x)Z_{QCD}[j,\bar\varepsilon,\varepsilon,\bar\eta_q,\eta_q]=\langle c^a(x)\rangle  \nonumber \\
		&&q_{1}^{(\eta)i}(x)Z_{QCD}[j,\bar\varepsilon,\varepsilon,\bar\eta_q,\eta_q]=\langle q^i(x)\rangle.
\end{eqnarray}
These represent the 1-point function for the gluon, ghost and quark fields respectively. Deriving once with respect to the currents, at the same point because of the averages on the vacuum (see \cite{Bender:1999ek}), one has
\begin{eqnarray}
   &&G_{2\nu\kappa}^{(j)ab}(x,x)Z_{YM}[j,\bar\varepsilon,\epsilon]+G_{1\nu}^{(j)a}(x)G_{1\kappa}^{(j)b}(x)Z_{YM}[j,\bar\varepsilon,\epsilon]=\langle A^a_\nu(x)A^b_\kappa(x)\rangle \nonumber \\
	 &&H^{(\varepsilon)ab}_2(x,x)Z_{YM}[j,\bar\varepsilon,\epsilon]+P^{(\varepsilon)a}_1(x)P^{(\varepsilon)b}_1(x)Z_{YM}[j,\bar\varepsilon,\epsilon]=\langle c^b(x)c^a(x)\rangle \nonumber \\
	 &&\bar H^{(\varepsilon)ab}_2(x,x)Z_{YM}[j,\bar\varepsilon,\epsilon]+\bar P^{(\varepsilon)a}_1(x)\bar P^{(\varepsilon)b}_1(x)Z_{YM}[j,\bar\varepsilon,\epsilon]=
	\langle \bar c^b(x)\bar c^a(x)\rangle \nonumber \\
	 &&P^{(\varepsilon)ab}_2(x,x)Z_{YM}[j,\bar\varepsilon,\epsilon]+\bar P^{(\varepsilon)a}_1(x)P^{(\varepsilon)b}_1(x)Z_{YM}[j,\bar\varepsilon,\epsilon]=
	\langle \bar c^b(x)c^a(x)\rangle \nonumber \\
	&&\partial_\mu G_{2\nu\kappa}^{(j)ab}(x,x)Z_{YM}[j,\bar\varepsilon,\epsilon]+\partial_\mu G_{1\nu}^{(j)a}(x)G_{1\kappa}^{(j)b}(x)Z_{YM}[j,\bar\varepsilon,\epsilon]=
	\langle\partial_\mu A^a_\nu(x)A^b_\kappa(x)\rangle \nonumber \\
	&&K^{(\varepsilon,j)ab}_{2\nu}(x,x)Z_{YM}[j,\bar\varepsilon,\epsilon]+P^{(\varepsilon)a}_1(x)G_{1\nu}^{(j)b}(x)Z_{YM}[j,\bar\varepsilon,\epsilon]=\langle c^a(x)A_\nu^b(x)\rangle \nonumber \\
	&&\slashed G_1^{(j)a}(x) q_{1}^{(\eta)i}(x)Z_{QCD}[j,\bar\varepsilon,\varepsilon,\bar\eta_q,\eta_q]+\slashed W^{(\eta)ai}_{q}(x,x)Z_{QCD}[j,\bar\varepsilon,\varepsilon,\bar\eta_q,\eta_q]=\langle\slashed A^a q^i(x)\rangle 
	\nonumber \\
	&& S^{(\eta)ij}_q(x,x)Z_{QCD}[j,\bar\varepsilon,\varepsilon,\bar\eta_q,\eta_q]+{\bar q}_1^{(\eta)i}(x)q_{1}^{(\eta)j}(x)Z_{QCD}[j,\bar\varepsilon,\varepsilon,\bar\eta_q,\eta_q]=
	\langle{\bar q}^i(x)q^j(x)\rangle \nonumber \\
	&& J_{q\mu}^{(\eta)aij}(x,x)Z_{QCD}[j,\bar\varepsilon,\varepsilon,\bar\eta_q,\eta_q]+{\bar q}_1^{(\eta)i}(x)\gamma_\mu T^a q_{1}^{(\eta)j}(x)
	Z_{QCD}[j,\bar\varepsilon,\varepsilon,\bar\eta_q,\eta_q]=
	\langle{\bar q}^i(x)\gamma_\mu T^a q^j(x)\rangle,
\end{eqnarray}
being $W^{(\eta)ai}_{q\mu}(x,y)=\delta G^{(j)a}_{1\mu}(x)/\delta \eta_q^i(y)$, $J_{q\mu}^{(\eta)aij}(x,y)= \gamma_\mu T^a\delta q_{1}^{(\eta)i}(x)/\delta\eta_q^j(y)=\gamma_\mu T^aS^{(\eta)ij}_q(x,y)$. Deriving twice one has
\begin{eqnarray}
   &&G_{3\nu\kappa\rho}^{(j)abc}(x,x,x)Z_{QCD}[j,\bar\varepsilon,\varepsilon,\bar\eta_q,\eta_q]+G_{2\nu\kappa}^{(j)ab}(x,x)G_{1\rho}^{(j)c}(x)Z_{QCD}[j,\bar\varepsilon,\varepsilon,\bar\eta_q,\eta_q]+ \nonumber \\
	&&G_{2\nu\rho}^{(j)ac}(x,x)G_{1\kappa}^{(j)b}(x)Z_{YM}[j,\bar\varepsilon,\epsilon]
	+G_{1\nu}^{(j)a}(x)G_{2\kappa\rho}^{(j)bc}(x,x)Z_{QCD}[j,\bar\varepsilon,\varepsilon,\bar\eta_q,\eta_q]+ \nonumber \\
	&&G_{1\nu}^{(j)a}(x)G_{1\kappa}^{(j)b}(x)G_{1\rho}^{(j)c}(x)Z_{QCD}[j,\bar\varepsilon,\varepsilon,\bar\eta_q,\eta_q]=\langle A^a_\nu(x)A^b_\kappa(x)A^c_\rho(x)\rangle.
\end{eqnarray}
This give us the first set of Schwinger-Dyson equations as
\begin{eqnarray}
\label{eq:ds_1}
    &&\partial^2G_{1\nu}^{(j)a}(x)+gf^{abc}(
		\partial^\mu G_{2\mu\nu}^{(j)bc}(x,x)+\partial^\mu G_{1\mu}^{(j)b}(x)G_{1\nu}^{(j)c}(x)-
		\partial_\nu G_{2\mu}^{(j)\mu bc}(x,x)-\partial_\nu G_{1\mu}^{(j)b}(x)G_{1}^{(j)\mu c}(x)) \nonumber \\
		&&+gf^{abc}\partial^\mu G_{2\mu\nu}^{(j)bc}(x,x)+gf^{abc}\partial^\mu(G_{1\mu}^{(j)b}(x)G_{1\nu}^{(j)c}(x))		
		\nonumber \\
		&&+g^2f^{abc}f^{cde}(G_{3\mu\nu}^{(j)\mu bde}(x,x,x)
		+G_{2\mu\nu}^{(j)bd}(x,x)G_{1}^{(j)\mu e}(x) \nonumber \\
	&&+G_{2\nu\rho}^{(j)eb}(x,x)G_{1}^{(j)\rho d}(x)
	+G_{2\mu\nu}^{(j)de}(x,x)G_{1}^{(j)\mu b}(x)+ \nonumber \\
	&&G_{1}^{(j)\mu b}(x)G_{1\mu}^{(j)d}(x)G_{1\nu}^{(j)e}(x))
		=gf^{abc}(\partial_\nu P^{(\varepsilon)bc}_2(x,x)+\partial_\nu (\bar P^{(\varepsilon)b}_1(x)P^{(\varepsilon)c}_1(x))) \nonumber \\
		&&+g\sum_{q,i}J_{q\nu}^{(\eta)aii}(x,x)+g\sum_{q,i}{\bar q}_1^{(\eta)i}(x)\gamma_\nu T^a q_{1}^{(\eta)i}(x)+ j_\nu^a \nonumber \\
	 &&\partial^2 P^{(\varepsilon)a}_1(x)+gf^{abc}\partial^\mu
	(K^{(\varepsilon,j)bc}_{2\mu}(x,x)+P^{(\varepsilon)b}_1(x)G_{1\mu}^{(j)c}(x))=\varepsilon^a \nonumber \\
	&& (i\slashed\partial-m_q)q_{1}^{(\eta)i}(x)+g{\bm T}\cdot\slashed{\bm G}_1^{(j)}(x) q_{1}^{(\eta)i}(x)+g{\bm T}\cdot\slashed{\bm W}^{(\eta)i}_q(x,x)q_{1}^{(\eta)i}(x) =\eta_q^i.
\end{eqnarray}
Setting the currents to zero and noticing that, by translation invariance, is $G_2(x,x)=G_2(x-x)=G_2(0)$, $G_3(x,x,x)=G_3(0,0)$, $K_2(x,x)=K_2(0)$, $W^{(\eta)a}(x,x)=W^{(\eta)a}(0)$ and $J_{q\nu}(x,x)=J_{q\nu}(0)$, we get
\begin{eqnarray}
\label{eq:ds_11}
    &&\partial^2G_{1\nu}^{a}(x)+gf^{abc}(
		\partial^\mu G_{2\mu\nu}^{bc}(0)+\partial^\mu G_{1\mu}^{b}(x)G_{1\nu}^{c}(x)-
		\partial_\nu G_{2\mu}^{\nu bc}(0)-\partial_\nu G_{1\mu}^{b}(x)G_{1}^{\mu c}(x)) \nonumber \\
		&&+gf^{abc}\partial^\mu G_{2\mu\nu}^{bc}(0)+gf^{abc}\partial^\mu(G_{1\mu}^{b}(x)G_{1\nu}^{c}(x))		
		\nonumber \\
		&&+g^2f^{abc}f^{cde}(G_{3\mu\nu}^{\mu bde}(0,0)
		+G_{2\mu\nu}^{bd}(0)G_{1}^{\mu e}(x) \nonumber \\
	&&+G_{2\nu\rho}^{eb}(0)G_{1}^{\rho d}(x)
	+G_{2\mu\nu}^{de}(0)G_{1}^{\mu b}(x)+ \nonumber \\
	&&G_{1}^{\mu b}(x)G_{1\mu}^{d}(x)G_{1\nu}^{e}(x))
		=gf^{abc}(\partial_\nu P^{bc}_2(0)+\partial_\nu (\bar P^{b}_1(x)P^{c}_1(x))) \nonumber \\
		&& g\sum_{q,i}\gamma_\nu T^aS_{q}^{ii}(0)+g\sum_{q,i}{\bar q}_1^i(x)\gamma_\nu T^a q_1^i(x) \nonumber \\
	 &&\partial^2 P^{a}_1(x)+gf^{abc}\partial^\mu
	(K^{bc}_{2\mu}(0)+P^{b}_1(x)G_{1\mu}^{c}(x))=0 \nonumber \\
	&&(i\slashed\partial-m_q)q_{1}^{i}(x)+g{\bm T}\cdot\slashed{\bm G}_1(x) q_{1}^{i}(x)+g{\bm T}\cdot\slashed{\bm W}^{i}_q(0)q_1^i(x) = 0.
\end{eqnarray}

The Schwinger-Dyson equation for the two-point functions can be obtained by further deriving eq.(\ref{eq:ds_1}). One has
\begin{eqnarray}
\label{eq:ds_2}
    &&\partial^2G_{2\nu\kappa}^{(j)ad}(x-y)
		+gf^{abc}(
		\partial^\mu G_{3\mu\nu\kappa}^{(j)bcd}(x,x,y)
		+\partial^\mu G_{2\mu\kappa}^{(j)bd}(x-y)G_{1\nu}^{(j)c}(x)
		+\partial^\mu G_{1\mu}^{(j)b}(x)G_{2\nu\kappa}^{(j)cd}(x-y) \nonumber \\
		&&-\partial_\nu G_{3\mu\kappa}^{(j)\mu bcd}(x,x,y)
		-\partial_\nu G_{2\mu\kappa}^{(j)bd}(x-y)G_{1}^{(j)\mu c}(x)) 
		-\partial_\nu G_{1\mu}^{(j)b}(x)G_{2\kappa}^{(j)\mu cd}(x-y))
		\nonumber \\
		&&+gf^{abc}\partial^\mu G_{3\mu\nu\kappa}^{(j)bcd}(x,x,y)
		+gf^{abc}\partial^\mu(G_{2\mu\kappa}^{(j)bd}(x-y)G_{1\nu}^{(j)c}(x))
				+gf^{abc}\partial^\mu(G_{1\mu}^{(j)b}(x)G_{1\nu\kappa}^{(j)cd}(x-y))
		\nonumber \\
		&&+g^2f^{abc}f^{cge}(G_{4\mu\nu\kappa}^{(j)\mu bged}(x,x,x,y)
		+G_{3\mu\nu\kappa}^{(j)bgd}(x,x,y)G_{1}^{(j)\mu e}(x) 
		+G_{2\mu\nu}^{(j)bg}(x,x)G_{2\kappa}^{(j)\mu ed}(x-y)\nonumber \\
	&&+G_{3\nu\rho\kappa}^{(j)acd}(x,x,y)G_{1}^{(j)\rho b}(x)
	+G_{2\nu\rho}^{(j)eb}(x,x)G_{2\kappa}^{(j)\rho gd}(x-y) \nonumber \\
	&&+G_{2\nu\rho}^{(j)ge}(x,x)G_{2\kappa}^{(j)\rho bd}(x-y)
	+G_{1}^{(j)\mu g}(x)G_{3\mu\nu\kappa}^{(j)ged}(x,x,y)+ \nonumber \\
	&&G_{2\kappa}^{(j)\mu bd}(x-y)G_{1\mu}^{(j)g}(x)G_{1\nu}^{(j)e}(x)+
	G_{1}^{(j)\mu b}(x)G_{2\mu\kappa}^{(j)gd}(x-y)G_{1\nu}^{(j)e}(x)+
	G_{1}^{(j)\mu b}(x)G_{1\mu}^{(j)g}(x)G_{2\nu\kappa}^{(j)ed}(x-y)) \nonumber \\
	&&	=gf^{abc}(\partial_\nu K^{(j\varepsilon)bcd}_{3\kappa}(x,x,y)
	+\partial_\nu (\bar P^{(\varepsilon)b}_1(x)K^{(j\varepsilon)cd}_{2\kappa}(x,y))) 
	+\partial_\nu (\bar K^{(j\varepsilon)bd}_{2\kappa}(x,y)P^{(\varepsilon)c}_1(x))) \nonumber \\
	&&+g\sum_{q,i}J_{q1\nu\kappa}^{(\eta)iiad}(0,x-y)
	+g\sum_{q,i}{\bar Q}^{(\eta)id}_\kappa(x-y)\gamma_\nu T^a q_{1}^{(\eta)i}(x) \nonumber \\
	&&+g\sum_{q,i}{\bar q}_1^{(\eta)i}(x)\gamma_\nu T^a Q^{(\eta)id}_\kappa(x-y)
	+ \delta_{ad}g_{\nu\kappa}\delta^4(x-y) \nonumber \\
	 &&\partial^2 P^{(\varepsilon)ad}_2(x-y)+gf^{abc}\partial^\mu
	(K^{(\varepsilon,j)bcd}_{3\mu}(x,x,y)+P^{(\varepsilon)bd}_2(x-y)G_{1\mu}^{(j)c}(x)+ \nonumber \\
	&&P^{(\varepsilon)b}_1(x)K_{2\mu}^{(j\varepsilon)cd}(x-y))=\delta_{ad}\delta^4(x-y) \nonumber \\
	&&\partial^2 K^{(j\varepsilon)ad}_{2\kappa}(x-y)+gf^{abc}\partial^\mu
	(L^{(\varepsilon,j)bcd}_{2\mu\kappa}(x,x,y)+ \nonumber \\
	&&K^{(j\varepsilon)bd}_{2\kappa}(x-y)G_{1\mu}^{(j)c}(x)+P^{(\varepsilon)b}_1(x)G_{2\mu\kappa}^{(j)cd}(x-y))=0 \nonumber \\
	&& (i\slashed\partial-m_q)S^{(\eta)ij}_q(x-y)+g{\bm T}\cdot\slashed {\bm G}^{(j)}_1(x) S^{(\eta)ij}_q(x-y)+g{\bm T}\cdot\slashed {\bm W}^{(\eta)j}_{1q}(x,x,y) q_{1}^{(\eta)i}(x) \nonumber \\
	&&+g{\bm T}\cdot\slashed{\bm W}^{(\eta)}(x,x)S^{(\eta)ij}_q(x-y) =\delta_{ij}\delta^4(x-y) \nonumber \\  
  &&\partial^2W_{q\nu}^{(\eta)ai}(x,y)+gf^{abc}(
		\partial^\mu W_{1q\mu\nu}^{(\eta)bci}(x,x,y)+\partial^\mu W_{q\mu}^{(\eta)bi}(x,y)G_{1\nu}^{(j)c}(x)
		+\partial^\mu G_{1\mu}^{(j)b}(x)W_{q\nu}^{(\eta)ci}(x,y) \nonumber \\
		&&-\partial_\nu W_{1q\mu}^{(\eta)\mu bci}(x,x,y)-\partial_\nu W_{q\mu}^{(\eta)bi}(x,y)G_{1}^{(j)\mu c}(x)
		-\partial_\nu G_{1\mu}^{(j)b}(x)W^{(\eta)\mu ci}_q(x,y)) \nonumber \\
  &&+gf^{abc}\partial^\mu W_{1q\mu\nu}^{(\eta)bci}(x,x,y)+gf^{abc}\partial^\mu(W_{q\mu}^{(\eta)bi}(x,y)G_{1\nu}^{(j)c}(x))	
		+gf^{abc}\partial^\mu(G_{1\mu}^{(j)b}(x)W_{q\nu}^{(\eta)ci}(x,y))
		\nonumber \\
	&&+g^2f^{abc}f^{cde}(W_{2q\mu\nu}^{(\eta)\mu bdei}(x,x,x,y)
		+W_{1q\mu\nu}^{(\eta)bdi}(x,x,y)G_{1}^{(j)\mu e}(x) 
		+G_{2\mu\nu}^{(j)bd}(x,x)W^{(\eta)\mu ei}_q(x,y)\nonumber \\
	&&+W_{1q\nu\rho}^{(\eta)ebi}(x,x,y)G_{1}^{(j)\rho d}(x)+G_{2\nu\rho}^{eb}(x,x)W^{(eta)\rho di}(x,y)
	+W_{1q\mu\nu}^{(eta)dei}(x,x,y)G_{1}^{\mu b}(x)
	+G_{2\mu\nu}^{(j)de}(x,x)W^{(\eta)\mu bi}_q(x,y)+
	\nonumber \\
	&&W^{(\eta)\mu bi}_q(x,y)G_{1\mu}^{(j)d}(x)G_{1\nu}^{(j)e}(x)+
	G_{1}^{(j)\mu b}(x)W_{q\mu}^{(\eta)di}(x,y)G_{1\nu}^{(j)e}(x)+
	G_{1}^{(j)\mu b}(x)G_{1\mu}^{(j)d}(x)W_{q\nu}^{(\eta)ei}(x,y)) \nonumber \\
	&&=gf^{abc}(\partial_\nu Y^{(\varepsilon)bci}_{1q}(x,x,y)+\partial_\nu (\bar Y^{(\varepsilon)bi}_q(x,y)P_1^{(\varepsilon)c}(x,y)+\bar P_1^{(\varepsilon)b}(x)Y^{(\varepsilon)ci}_q(x))) \nonumber \\
	&&+g\sum_{j}J_{1q\nu}^{(\eta)ajji}(x,x,y)+g\sum_{j}{\bar q}_1^{(\eta)j}(x)\gamma_\nu T^a S^{(\eta)ji}_q(x,y) \nonumber \\
	&& (i\slashed\partial-m_q)Q^{(\eta)ia}_\mu(x,y)+g{\bm T}\cdot\slashed{\bm G}_{2\mu}^{(j)a}(x,y) q_{1}^{(\eta)i}(x)
	+g{\bm T}\cdot\slashed{\bm G}_{1}^{(j)}(x) Q^{(\eta)ia}_\mu(x,y) \nonumber \\
	&&+g{\bm T}\cdot\slashed{\bm Z}^{(\eta)ia}_{q\mu}(x,x,y)q_{1}^{(\eta)i}(x)+g{\bm T}\cdot\slashed{\bm W}^{(\eta)i}_q(x,x)Q^{(\eta)ia}_\mu(x,y) = 0.
\end{eqnarray}
We have assumed $\langle\bar q\bar q\rangle=0$, $\langle \bar q' q\rangle=0$ for $q'\ne q$, $Y^{(\varepsilon)b\ldots i}_{nq}(x,y,\ldots,z)=\delta P_{n+1}^{(\varepsilon)b\ldots}(x,y,\ldots)/\delta\eta_q^i$, $Q^{(\eta)ia}_\mu(x-y)=\delta q_1^{(eta)i}(x)/\delta j^a_\mu(y)$, $J_{1q\mu}^{(\eta)aijk}(x,y,z)=\delta J_{q\mu}^{(\eta)aij}(x,y)/\delta\eta^k_q(z)=\gamma_\mu T_a\delta S^{(\eta)ij}_q(x,y)/\delta\eta^k_q(z)$, $W^{(\eta)a\ldots i}_{nq\mu}(x,y,\ldots,z)=\delta G^{(j)a\ldots}_{{n+1}\mu}(x,y,\ldots)/\delta{\eta}_q^j(z)$. Note that, for $n=0$, we omit the index, e.g. $W_0=W$. This yields, setting currents to zero and using translation invariance,
\begin{eqnarray}
\label{eq:ds_22}
    &&\partial^2G_{2\nu\kappa}^{ad}(x-y)+gf^{abc}(
		\partial^\mu G_{3\mu\nu\kappa}^{bcd}(0,x-y)+\partial^\mu G_{2\mu\kappa}^{bd}(x-y)G_{1\nu}^{c}(x)
		+\partial^\mu G_{1\mu}^{b}(x)G_{2\nu\kappa}^{cd}(x-y) \nonumber \\
		&&-\partial_\nu G_{3\mu\kappa}^{\mu bcd}(0,x-y)-\partial_\nu G_{2\mu\kappa}^{bd}(x-y)G_{1}^{\mu c}(x)) 
		-\partial_\nu G_{1\mu}^{b}(x)G_{2\kappa}^{\mu cd}(x-y))
		\nonumber \\
		&&+gf^{abc}\partial^\mu G_{3\mu\nu\kappa}^{bcd}(0,x-y)
		+gf^{abc}\partial^\mu(G_{2\mu\kappa}^{bd}(x-y)G_{1\nu}^{c}(x))
				+gf^{abc}\partial^\mu(G_{1\mu}^{b}(x)G_{1\nu\kappa}^{cd}(x-y))
		\nonumber \\
		&&+g^2f^{abc}f^{cge}(G_{4\mu\nu\kappa}^{\mu bged}(0,0,x-y)
		+G_{3\mu\nu\kappa}^{bgd}(0,x-y)G_{1}^{\mu e}(x) 
		+G_{2\mu\nu}^{bg}(0)G_{2\kappa}^{\mu ed}(x-y)\nonumber \\
	&&+G_{3\nu\rho\kappa}^{acd}(0,x-y)G_{1}^{\rho b}(x)
	+G_{2\nu\rho}^{eb}(0)G_{2\kappa}^{\rho gd}(x-y)
	+G_{2\nu\rho}^{ge}(0)G_{2\kappa}^{\rho bd}(x-y)
	+G_{1}^{\mu b}(x)G_{3\mu\nu\kappa}^{ged}(0,x-y)+ \nonumber \\
	&&G_{2\kappa}^{\mu bg}(x-y)G_{1\mu}^{d}(x)G_{1\nu}^{e}(x)+
	G_{1}^{\mu b}(x)G_{2\mu\kappa}^{gd}(x-y)G_{1\nu}^{e}(x)+
	G_{1}^{\mu b}(x)G_{1\mu}^{g}(x)G_{2\nu\kappa}^{ed}(x-y)) \nonumber \\
	&&=gf^{abc}(\partial_\nu K^{bcd}_{3\kappa}(0,x-y)+\partial_\nu (\bar P^{b}_1(x)K^{cd}_{2\kappa}(x-y))) 
	+\partial_\nu (\bar K^{bd}_{2\kappa}(x-y)P^{c}_1(x))) \nonumber \\	
	&&+g\sum_{q,i}J_{q1\nu\kappa}^{iiad}(0,x-y)
	+g\sum_{q,i}{\bar Q}^{id}_\kappa(x-y)\gamma_\nu T^a q_{1}^{i}(x)
	+g\sum_{q,i}{\bar q}_1^{i}(x)\gamma_\nu T^a Q^{id}_\kappa(x-y)
	+ \delta_{ad}g_{\nu\kappa}\delta^4(x-y) \nonumber \\
	 &&\partial^2 P^{ad}_2(x-y)+gf^{abc}\partial^\mu
	(K^{bcd}_{3\mu}(0,x-y)+P^{bd}_2(x-y)G_{1\mu}^{c}(x)+ \nonumber \\
	&&P^{b}_1(x)K_{2\mu}^{cd}(x-y))=\delta_{ad}\delta^4(x-y) \nonumber \\
	&&\partial^2 K^{ad}_{2\kappa}(x-y)+gf^{abc}\partial^\mu
	(L^{bcd}_{2\mu\kappa}(0,x-y)+ \nonumber \\
	&&K^{bd}_{2\kappa}(x-y)G_{1\mu}^{c}(x)+P^{b}_1(x)G_{2\mu\kappa}^{cd}(x-y))=0 \nonumber \\
	&& (i\slashed\partial-m_q)S^{ij}_q(x-y)+g{\bm T}\cdot\slashed {\bm G}_1(x) S^{ij}_q(x-y)+g{\bm T}\cdot\slashed {\bm W}^{j}_{1q}(0,x-y) q_{1}^{i}(x) \nonumber \\
	&&+g{\bm T}\cdot\slashed{\bm W}(0)S^{ij}_q(x-y) =\delta_{ij}\delta^4(x-y) \nonumber \\
  &&\partial^2W_{q\nu}^{ai}(x,y)+gf^{abc}(
		\partial^\mu W_{1q\mu\nu}^{bci}(0,x-y)+\partial^\mu W_{q\mu}^{bi}(x,y)G_{1\nu}^{c}(x)
		+\partial^\mu G_{1\mu}^{b}(x)W_{q\nu}^{ci}(x,y)
		-\partial_\nu W_{1q\mu}^{\mu bci}(0,x-y) \nonumber \\
  &&-\partial_\nu W_{q\mu}^{bi}(x,y)G_{1}^{\mu c}(x)
		-\partial_\nu G_{1\mu}^{b}(x)W^{\mu ci}_q(x,y)) \nonumber \\
  &&+gf^{abc}\partial^\mu W_{1q\mu\nu}^{bci}(0,x-y)+gf^{abc}\partial^\mu(W_{q\mu}^{bi}(x,y)G_{1\nu}^{c}(x))	
		+gf^{abc}\partial^\mu(G_{1\mu}^{b}(x)W_{q\nu}^{ci}(x,y))
		\nonumber \\
	&&+g^2f^{abc}f^{cde}(W_{2q\mu\nu}^{\mu bdei}(0,0,x-y)
		+W_{1q\mu\nu}^{bdi}(0,x-y)G_{1}^{\mu e}(x) 
		+G_{2\mu\nu}^{bd}(0)W^{\mu ei}_q(x,y)\nonumber \\
	&&+W_{1q\nu\rho}^{ebi}(0,x-y)G_{1}^{\rho d}(x)+G_{2\nu\rho}^{eb}(0)W^{\rho di}_q(x,y)
	+G_{2\mu\nu}^{de}(0)W^{\mu bi}_q(x,y)+W_{1q\mu\nu}^{dei}(0,x-y)G_{1}^{\mu b}(x)+
	\nonumber \\
	&&W^{\mu bi}_q(x,y)G_{1\mu}^{d}(x)G_{1\nu}^{e}(x)+
	G_{1}^{\mu b}(x)W_q{\mu}^{di}(x,y)G_{1\nu}^{e}(x)+
	G_{1}^{\mu b}(x)G_{1\mu}^{d}(x)W_{q\nu}^{ei}(x,y)) \nonumber \\
	&&=gf^{abc}(\partial_\nu Y^{bci}_{1q}(0,x-y)+\partial_\nu (\bar Y^{bi}_q(x,y)P_1^{c}(x,y)+\bar P_1^{b}(x)Y^{ci}_q(x,y))) \nonumber \\
	&&+g\sum_{j}J_{1q\nu}^{ajji}(0,x-y)+g\sum_{j}{\bar q}_1^{j}(x)\gamma_\nu T^a S^{ji}_q(x,y)
\end{eqnarray}
\begin{eqnarray}
	(i\slashed\partial-m_q)Q^{ia}_\mu(x,y)+g{\bm T}\cdot\slashed{\bm G}_{2\mu}^{(j)a}(x,y) q_{1}^{i}(x)
	+g{\bm T}\cdot\slashed{\bm G}_{1}(x) Q^{ia}_\mu(x,y)&& \nonumber \\
	+g{\bm T}\cdot\slashed{\bm Z}^{ia}_{q\mu}(0,x-y)q_{1}^{i}(x)+g{\bm T}\cdot\slashed{\bm W}^{i}_q(0)Q^{ia}_\mu(x,y) &=& 0. \nonumber
\end{eqnarray}

\section{'t~Hooft limit\label{sec3}}

We will give here an approximate solution to the set of equations we obtained in the preceding section.
The technique we will use is an iterative one starting from the exact solutions we obtained or the 1- and 2-point functions of the Yang-Mills theory  \cite{Frasca:2015yva}. These correlation functions will be modified by the presence of quarks but, in a first iteration, we assume they are a good non-perturbative approximation working well in the deep infrared limit. In this way, we will be able to derive a confining Nambu-Jona-Lasinio approximation.
We start from the set (\ref{eq:ds_11}) and take
\begin{equation}
P^{c}_1(x)=0, \ {\bar P}^{c}_1(x)=0.
\end{equation}
For symmetry properties of $P_2$ under exchange of indexes the set reduces to
\begin{eqnarray}
\label{eq:ds_110}
    &&\partial^2G_{1\nu}^{a}(x)+gf^{abc}(
		\partial^\mu G_{2\mu\nu}^{bc}(0)+\partial^\mu G_{1\mu}^{b}(x)G_{1\nu}^{c}(x)-
		\partial_\nu G_{2\mu}^{\nu bc}(0)-\partial_\nu G_{1\mu}^{b}(x)G_{1}^{\mu c}(x)) \nonumber \\
		&&+gf^{abc}\partial^\mu G_{2\mu\nu}^{bc}(0)+gf^{abc}\partial^\mu(G_{1\mu}^{b}(x)G_{1\nu}^{c}(x))		
		\nonumber \\
		&&+g^2f^{abc}f^{cde}(G_{3\mu\nu}^{\mu bde}(0,0)
		+G_{2\mu\nu}^{bd}(0)G_{1}^{\mu e}(x) \nonumber \\
	&&+G_{2\nu\rho}^{eb}(0)G_{1}^{\rho d}(x)
	+G_{2\mu\nu}^{de}(0)G_{1}^{\mu b}(x)+ \nonumber \\
	&&G_{1}^{\mu b}(x)G_{1\mu}^{d}(x)G_{1\nu}^{e}(x))
		=g\sum_{q,i}\gamma_\nu T^aS_{q}^{ii}(0)+g\sum_{q,i}{\bar q}_1^i(x)\gamma_\nu T^a q_1^i(x) \nonumber \\
	&&(i\slashed\partial-{\hat M}_q)q_{1}^{i}(x)+g{\bm T}\cdot\slashed{\bm G}_1(x) q_{1}^{i}(x) = 0.
\end{eqnarray}
We have introduced the quark mass matrix
\begin{equation}
{\hat M}_q^i=m_qI-g{\bm T}\cdot\slashed{\bm W}^{i}_q(x,x).
\end{equation} 
We note that is degenerate with respect the color index. Now, we write the expected solution in the form in the Landau gauge
\begin{eqnarray}
\label{eq:12p}
G_{1\nu}^a(x)&=&\eta_\nu^a\phi(x) \nonumber \\
G_{2\mu\nu}^{bd}(x-y)&=&\left(g_{\mu\nu}-\frac{\partial_\mu\partial_\nu}{\partial^2}\right)\delta^{bd}\Delta(x-y)
\end{eqnarray}
being $\eta_\nu^a$ some numerical coefficients, $\phi(x)$ a scalar field and $\Delta(x-y)$ the propagator. We will get for the 1-point functions
\begin{eqnarray}
\label{eq:ds_111}
  &&\eta_\nu^a\partial^2\phi(x)+2Ng^2\Delta(0)\eta_\nu^a\phi(x)+Ng^2\eta_\nu^a\phi^3(x) \nonumber \\
	&&=g\sum_{q,i}\gamma_\nu T^aS_{q}^{ii}(0)+g\sum_{q,i}{\bar q}_1^i(x)\gamma_\nu T^a q_1^i(x) \nonumber \\
	&&(i\slashed\partial-{\hat M}_q^i)q_{1}^{i}(x)+g{\bm T}\cdot\slashed\eta\phi(x) q_{1}^{i}(x) = 0.
\end{eqnarray}
In order to understand the behavior of the $\eta-$symbols, we work out the case for SU(2) where the proof of their existence is straightforward. Indeed, in this case, they can be defined as
\begin{equation}
\eta_\mu^a=((0,1,0,0),(0,0,1,0),(0,0,0,1)),
\end{equation}
that yields
\begin{equation}
\eta_\mu^1=(0,1,0,0),\ \eta_\mu^2=(0,0,1,0),\ \eta_\mu^3=(0,0,0,1),
\end{equation}
that implies $\eta_\mu^a\eta^{a\mu}=3$. This generalizes to SU(N) as 
\begin{equation}
\label{eq:eta1}
\eta_\mu^a\eta^{a\mu}=N^2-1.
\end{equation}
Similarly, by generalizing the SU(2) case,
\begin{equation}
\eta_\mu^a\eta^{b\mu}=\delta_{ab},
\end{equation}
and
\begin{equation}
\eta_\mu^a\eta_\nu^a=\frac{1}{2}\left(g_{\mu\nu}-\delta_{\mu\nu}\right),
\end{equation}
being $g_{\mu\nu}$ the Minkowski metric and $\delta_{\mu\nu}$ the identity tensor. We note that the number of components of $\eta_\mu^a$ are identical to those of $\gamma_\mu T^a$, as can be checked already for the aforementioned SU(2) case, as expected and the equation is consistent. This permits to write for the 1-point functions, using eq.~(\ref{eq:eta1}),
\begin{eqnarray}
\label{eq:ds_112}
  &&\partial^2\phi(x)+2Ng^2\Delta(0)\phi(x)+Ng^2\phi^3(x) \nonumber \\
	&&=\frac{1}{N^2-1}\left[g\sum_{q,i}\eta^{a\nu}\gamma_\nu T^aS_{q}^{ii}(0)+g\sum_{q,i}{\bar q}_1^i(x)\eta^{a\nu}\gamma_\nu T^a q_1^i(x)\right] \nonumber \\
	&&(i\slashed\partial-{\hat M}_q^i)q_{1}^{i}(x)+g{\bm T}\cdot\slashed\eta\phi(x) q_{1}^{i}(x) = 0,
\end{eqnarray}
where, in the first equation, a summation on the repeated index $a$ is implied giving explicitly the term $\eta\cdot\gamma\cdot T$. Therefore, the 1-point function acquires a mass term given by $\mu_0^2=Ng^2\Delta(0)$. Now, we apply eq.~(\ref{eq:12p}) to the 2-point function set to obtain
\begin{eqnarray}
\label{eq:ds_220}
  &&\partial^2\Delta(x-y)+2Ng^2\Delta(0)\Delta(x-y)+3Ng^2\phi^2(x)\Delta(x-y) \nonumber \\
	&&=g\sum_{q,i}{\bar Q}^{ia}_\nu(x-y)\gamma^\nu T^a q_{1}^{i}(x)
	+g\sum_{q,i}{\bar q}_1^{i}(x)\gamma^\nu T^a Q^{ia}_\nu(x-y) + \delta^4(x-y) \nonumber \\
	&&\partial^2 P^{ad}_2(x-y)=\delta_{ad}\delta^4(x-y) \nonumber \\
	&& (i\slashed\partial-{\hat M}_q^i)S^{ij}_q(x-y)+g{\bm T}\cdot\slashed\eta\phi(x) S^{ij}_q(x-y)=\delta_{ij}\delta^4(x-y) \nonumber \\  
  &&\partial^2W_{q\nu}^{ai}(x-y)+2Ng^2\Delta(0)W_{q\nu}^{ai}(x-y)+3Ng^2\phi^2(x)W_{q\nu}^{ai}\nonumber \\
	&&=g\sum_{j}{\bar q}_1^{j}(x)\gamma_\nu T^a S^{ji}_q(x-y) \nonumber \\
	&& (i\slashed\partial-{\hat M}_q^i)Q^{ia}_\mu(x-y)
	+g{\bm T}\cdot\slashed\eta\phi(x) Q^{ia}_\mu(x-y)+gT^a\gamma_\mu\Delta(x-y) q_{1}^{i}(x)=0. 
\end{eqnarray}

This set of equations can be solved in the 't~Hooft limit $N\gg 1$, keeping $Ng^2$ constant, and $Ng^2\gg 1$. 
We have to show the ordering in $g$ for these equations, as by now we are applying a perturbation technique for a strongly coupled theory. This can be accomplished by a perturbation theory devised in \cite{Frasca:2013tma}. In order to get the right ordering of terms in a strong coupling limit, we rescale the space-time variables as $x_\mu\rightarrow \sqrt{Ng^2}x_\mu$. E.g. this implies for the first equation in (\ref{eq:ds_112})
\begin{eqnarray}
  \partial^2\phi(x')+2\Delta(0)\phi(x')+3\phi^3(x')&=& \nonumber \\
\frac{1}{\sqrt{Ng^2}\sqrt{N}(N^2-1)}\left[\sum_{q,i}\eta\cdot\gamma\cdot TS_{q}^{ii}(0)+\sum_{q,i}{\bar q}_1^i(x')\eta\cdot\gamma\cdot T q_1^i(x')\right].&&  
\end{eqnarray}Therefore, we realize that the RHS is at least of order $O(1/\sqrt{Ng^2})$ and so, negligible with respect to the LHS in the 't~Hooft limit as the scalar field scales accordingly. A similar argument can be applied to eqs.~(\ref{eq:ds_220}).
For the quark equation, we will show in a moment that the $\phi$ field, in this approximation, will take a factor $1/\sqrt{Ng^2}$. Essentially, the idea, for the gluon field, is that it is of order $Ng^2$ while the corrections implied by the quark field are always $\sqrt{Ng^2}$ implying the start of an iterative procedure to solve this system of equations. 
Then, our leading order equations can be written as
\begin{eqnarray}
\label{eq:ds_1121}
  &&\partial^2\phi_0(x)+2Ng^2\Delta(0)\phi_0(x)+3Ng^2\phi_0^3(x)=0 \nonumber \\
	&&(i\slashed\partial-{\hat M}_q^i){\hat q}_{1}^{i}(x)= 0.
\end{eqnarray}
Indeed, for 1-point function we have the non-trivial solution \cite{Frasca:2017slg}
\begin{equation}
\phi_0(x)=\sqrt{\frac{2\mu^4}{m^2+\sqrt{m^4+2Ng^2\mu^4}}}{\rm sn}\left(p\cdot x+\chi,\kappa\right)
\end{equation}
being $\mu$ and $\chi$ arbitrary integration constants and $\kappa=\frac{-m^2+\sqrt{m^4+2Ng^2\mu^4}}{-m^2-\sqrt{m^4+2Ng^2\mu^4}}$. We have set $m^2=2Ng^2\Delta(0)$ and taken the momenta $p$ so that
\begin{equation}
\label{eq:disp}
    p^2=m^2+\frac{Ng^2\mu^4}{m^2+\sqrt{m^4+2Ng^2\mu^4}}.
\end{equation}
In this way, we are able to see the consistency of our approximation as $\phi\sim O(1/\sqrt{Ng^2})$ and then, when applied to the LHS of the given equations, this changes them in higher order contributions. We just observe that the leading order is non-perturbative, due to its dependency on the coupling, and non-trivial and so, it is consistent in the infrared limit.
For this reason, we write the second set of equations as
\begin{eqnarray}
\label{eq:ds_222}
  &&\partial^2\Delta(x,y)+2Ng^2\Delta(0)\Delta(x-y)+3Ng^2\phi^2(x)\Delta(x-y) \nonumber \\
	&&=g\sum_{q,i}{\bar Q}^{ia}_\nu(x,y)\gamma^\nu T^a {\hat q}_{1}^{i}(x)
	+g\sum_{q,i}{\bar{\hat q}}_1^{i}(x)\gamma^\nu T^a Q^{ia}_\nu(x,y) + \delta^4(x-y) \nonumber \\
	&&\partial^2 P^{ad}_2(x-y)=\delta_{ad}\delta^4(x-y) \nonumber \\
	&& (i\slashed\partial-{\hat M}_q^i){\hat S}^{ij}_q(x-y)=\delta_{ij}\delta^4(x-y) \nonumber \\  
  &&\partial^2W_{q\nu}^{ai}(x,y)+2Ng^2\Delta(0)W_{q\nu}^{ai}(x,y)+3Ng^2\phi^2(x)W_{q\nu}^{ai}(x,y)\nonumber \\
	&&=g\sum_{j}{\bar {\hat q}}_1^{j}(x)\gamma_\nu T^a {\hat S}^{ji}(x-y) \nonumber \\
	&& (i\slashed\partial-{\hat M}_q^i){\hat Q}^{ia}_\mu(x,y)+gT^a\gamma_\mu\Delta(x-y) {\hat q}_{1}^{i}(x)=0. 
\end{eqnarray}
This set of equations can be solved exactly and yields the self-consistency equations of the theory. In order to present the solutions, we point out that the equation
\begin{equation}
\partial^2\Delta_0(x-y)+[m^2+3Ng^2\phi_0^2(x)]\Delta_0(x-y)=\delta^4(x-y)
\end{equation}
has the solution
\begin{equation}
\label{eq:propg}
   \Delta_0(p)=M{\hat Z}(\mu,m,Ng^2)\frac{2\pi^3}{K^3(\kappa)}
	\sum_{n=0}^\infty(-1)^n\frac{e^{-(n+\frac{1}{2})\pi\frac{K'(\kappa)}{K(\kappa)}}}{1-e^{-(2n+1)\frac{K'(\kappa)}{K(\kappa)}\pi}}(2n+1)^2\frac{1}{p^2-m_n^2+i\epsilon}
\end{equation}
Here we have set
\begin{equation}
M=\sqrt{m^2+\frac{Ng^2\mu^4}{m^2+\sqrt{m^4+2Ng^2\mu^4}}}, 
\end{equation}
and
\begin{equation}
{\hat Z}(\mu,m,Ng^2) = \frac{1}{4}
\frac
{(m^2+\sqrt{m^4+2Ng^2\mu^4})^\frac{7}{2}[N^2g^4\mu^8+2m^2(m^2+\sqrt{m^4+2\lambda\mu^4})(m^4+2Ng^2\mu^4)]}
{\splitfrac{\sqrt{m^4+2Ng^2\mu^4+m^2\sqrt{m^4+2Ng^2\mu^4}}(6N^2g^4\mu^{12}m^2+N^3g^6\mu^{12}\sqrt{m^4+2Ng^2\mu^4}}{\splitfrac{+19N^2g^4\mu^8m^6+9N^2g^4\mu^8m^4\sqrt{m^4+2Ng^2\mu^4}
+16Ng^2\mu^4m^{10}}{+2Ng^2\mu^4m^8\sqrt{m^4+2Ng^2\mu^4}+4m^{12}\sqrt{m^4+2Ng^2\mu^4})}}}.
\end{equation}
The mass spectrum is given by
\begin{equation}
\label{eq:spec}
   m_n=(2n+1)\frac{\pi}{2K(\kappa)}\sqrt{m^2+\frac{Ng^2\mu^4}{m^2+\sqrt{m^4+2Ng^2\mu^4}}}.
\end{equation}
Then,
\begin{eqnarray}
m^2&=&2Ng^2 M{\hat Z}(\mu,m,Ng^2)\frac{2\pi^3}{K^3(\kappa)}
	\sum_{n=0}^\infty(-1)^n\frac{e^{-(n+\frac{1}{2})\pi\frac{K'(\kappa)}{K(\kappa)}}}{1-e^{-(2n+1)\frac{K'(\kappa)}{K(\kappa)}\pi}}(2n+1)^2\times \nonumber \\
	&&\int\frac{d^dp}{(2\pi)^d}\frac{1}{p^2-(2n+1)^2M^2+i\epsilon}.
\end{eqnarray}
This self-consistency equation provides the proper spectrum of a Yang-Mills theory with no fermions \cite{Frasca:2017slg}, in very close agreement with lattice data. Therefore, we get immediately
\begin{eqnarray}
\label{eq:sols}
{\hat S}^{ij}_q(x,y)&=&\delta_{ij}(i\slashed\partial-{\hat M}_q^i)^{-1}\delta^4(x-y)\nonumber \\
{\hat Q}^{ia}_\mu(x,y)&=&-g\int d^4y'\sum_j{\hat S}^{ij}_q(x-y')T^a\gamma_\mu\Delta(y',y) {\hat q}_{1}^{j}(y') \nonumber \\
W_{q\nu}^{ai}(x,y)&=&g\int d^4y'\Delta_0(x-y')\sum_{j}{\bar {\hat q}}_1^{j}(y')\gamma_\nu T^a {\hat S}^{ji}_q(y'-y) \nonumber \\
\Delta(x,y)&=&\Delta_0(x-y)+g\int d^4y'\Delta_0(x-y')\left[\sum_{q,i}{\bar {\hat Q}}^{ia}_\nu(y',y)\gamma^\nu T^a {\hat q}_{1}^{i}(y')\right. \nonumber \\
	&&\left.+\sum_{q,i}{\bar{\hat q}}_1^{i}(y')\gamma^\nu T^a {\hat Q}^{ia}_\nu(y',y)\right].
\end{eqnarray}
The last equation is an integral equation to be solved iteratively. This will yield the corrections to the gluon propagator. 

Similarly, for the quark mass we have to solve the self-consistent system of equations
\begin{eqnarray}
\label{eq:qprop}
&&(i\slashed\partial-{\hat M}_q^i){\hat q}_{1}^{i}(x)= 0 \nonumber \\
&& (i\slashed\partial-{\hat M}_q^i){\hat S}^{ij}_q(x-y)=\delta_{ij}\delta^4(x-y),
\end{eqnarray}
given the mass matrix
\begin{equation}
\label{eq:M_q}
{\hat M}_q^i=m_qI-g^2\int d^4y'\Delta_0(x-y')T^a\gamma^\nu\sum_{k}{\bar {\hat q}}_1^{k}(y')\gamma_\nu T^a {\hat S}^{ki}_q(y'-x)
\end{equation}
that couples both. This set will yield the quark propagator. These equations can be written as
\begin{eqnarray}
&&(i\slashed\partial-m_q){\hat q}_{1}^{i}(x)=-g^2\int d^4y'\Delta_0(x-y')T^a\gamma^\nu\sum_{j}{\bar {\hat q}}_1^{j}(y')\gamma_\nu T^a {\hat S}^{ji}_q(y'-x){\hat q}_{1}^{i}(x)\\
&& (i\slashed\partial-m_q){\hat S}^{ij}_q(x-y)=\delta_{ij}\delta^4(x-y)
-g^2\int d^4y'\Delta_0(x-y')T^a\gamma^\nu\sum_{k}{\bar {\hat q}}_1^{k}(y')\gamma_\nu T^a {\hat S}^{ki}_q(y'-x){\hat S}^{ij}_q(x-y) \nonumber
\end{eqnarray}
At this stage, we can act perturbatively by iterating on the leading order solutions. We will see that the solution is consistent as it should. Then,
the
first approximation for ${\hat q}_{1}^{i}(x)=q_0^i(x)$ is the free particle solution. The second equation will yield the approximation
\begin{equation}
\Sigma^i_q(x,x)=g^2\int d^4y'\Delta_0(x-y')T^a\gamma^\nu\sum_{j}{\bar {\hat q}}_0^{j}(y')\gamma_\nu T^a {\hat S}^{ji}_{0q}(y'-x)
\end{equation}
that is a non-local Nambu-Jona-Lasinio approximation provided we identify ${\hat S}^{ji}_{0q}(y'-x)$ with the free Dirac propagator.


%
%
\section{Dynamical chiral symmetry breaking and confinement in the 't~Hooft limit\label{sec4a}}
%

\begin{figure}[H]
  \includegraphics[width=1\textwidth]{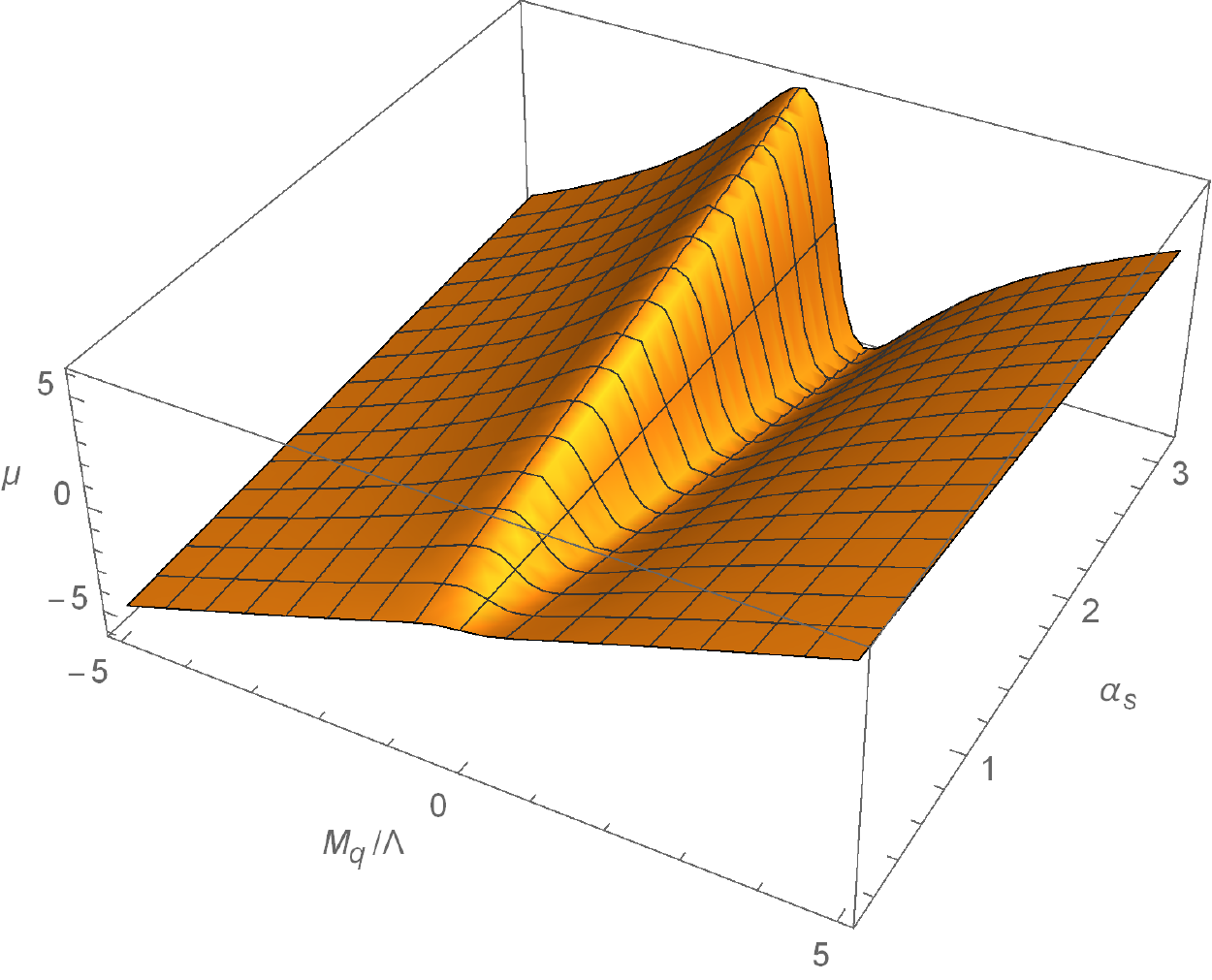}
  \caption{Equation of the effective mass, $\mu(x,y)$, as a function of $M_q/\Lambda$ and $\alpha_s$ and fixing $m_0=0.417$. Here $N=3$, $N_f=6$, $m_q=m_u=2\ {\rm MeV}$ and $\Lambda=1\ {\rm GeV}$. \label{fig:Meff}}
\end{figure}

\begin{figure}[H]
  \includegraphics[width=1\textwidth]{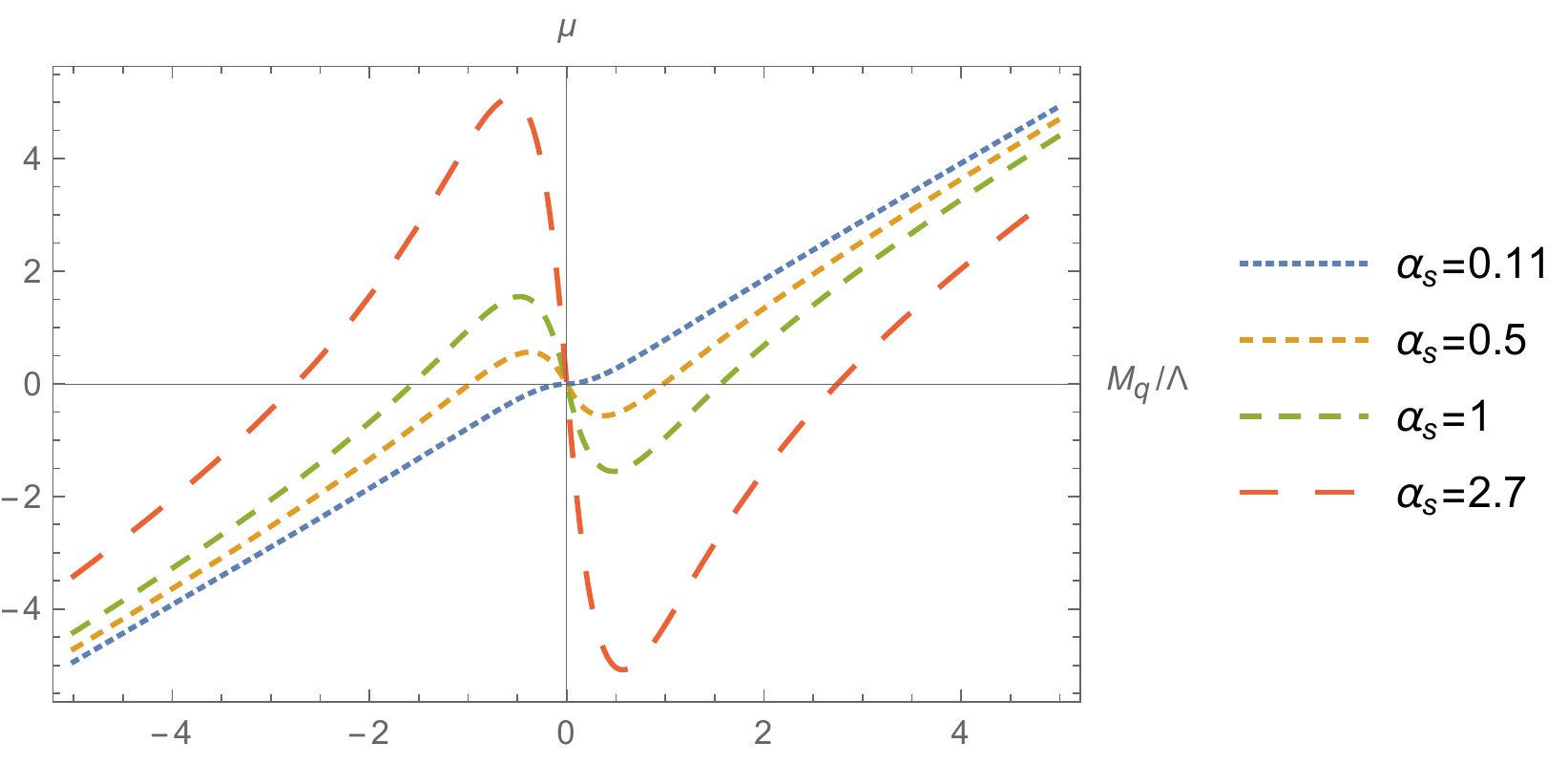}
  \caption{Section curves of $\mu(x,y)$, as a function of $M_q/\Lambda$, for different values of $\alpha_s$. We consider unphysical the region with $M_q/\Lambda>1$. \label{fig:Meff2}}
\end{figure}

%
%
We now discuss how dynamical chiral symmetry breaking arises naturally in the 't ~Hooft limit giving also
a condition for
 confinement of quarks. In order to do this we need to evaluate the quark propagator. We consider eq.(\ref{eq:qprop}) and we take the Nambu-Jona-Lasinio approximation. One has
\begin{equation}
(i\slashed\partial-m_q+\Sigma^i_{NJL}(x,x)){\hat S}^{ij}_q(x-y)=\delta_{ij}\delta^4(x-y)
\end{equation}
being 
the quark self-energy
\begin{equation}
\Sigma^i_{NJL}(x,x)=g^2\int d^4y'\Delta_0(x-y')T^a\gamma^\nu\sum_{j}{\bar {\hat q}}_0^{j}(y')\gamma_\nu T^a {\hat S}^{ji}_{0q}(y'-x).
\end{equation}
We realize that the non-local kernel comes out from the gluon propagator $\Delta_0(x-y)$. We note that
\begin{equation}
\label{eq:qself}
\Sigma^i_{NJL}(p)=g^2\int\frac{d^4p_1}{(2\pi)^4}\Delta_0(p_1)T^a\gamma^\nu\sum_{j}{\bar {\hat q}}_0^{j}(p)\gamma_\nu T^a {\hat S}^{ji}_{0q}(p_1-p).
\end{equation}
Quark self-energy says us that the corrections to the quark propagator can depend on momenta. So, we will have dynamical symmetry breaking until a solution for the quark mass can be found. When the on-shell mass condition fails, varying the coupling, we will have a confined quark. This is a possible definition of quark confinement given by Gribov \cite{Gribov:1998kb} and Roberts and Williams \cite{Roberts:1994dr}.
%
Therefore, we compute the gap equation in the limit of very low momenta. This will yield
\begin{equation}
M_q=m_q-{\rm Tr}\Sigma^i_{NJL}(0)
\end{equation}
where the trace is over flavors, colors and spinor indexes. This yields
\begin{equation}
M_q=m_q+\frac{N_f(N^2-1)Ng^2}{2}\int\frac{d^4p}{(2\pi)^4}\Delta_0(p)\frac{M_q}{p^2+M^2_q}
\end{equation}
having moved to Euclidean. This equation admits a real solution for the local limit $\Delta_0(p)\rightarrow{\rm constant}$ for a critical value of the coupling $g$. This implies a free quark and so, no confinement. In the general case, from the gluon propagator eq.~(\ref{eq:propg}), we can write (in the Euclidean limit)
\begin{equation}
\Delta_0(p)=\sum_{n=0}^\infty\frac{B_n}{p^2+m_n^2}
\end{equation}
with the mass spectrum, $m_n$, given in eq.~(\ref{eq:spec}). We assume no degeneracy between $m_n$ and $M_q$ and we approximate the gluon propagator in eq.(\ref{eq:propg}) with the correction to the mass $m^2$ neglected. This is a fairly good approximation as shown in \cite{Frasca:2017slg}. So, we have to evaluate
\begin{equation}
M_q=m_q+\frac{N_f(N^2-1)Ng^2}{2}\int\frac{d^4p}{(2\pi)^4}\sum_{n=0}^\infty\frac{B_n}{p^2+m_n^2}\frac{M_q}{p^2+M^2_q}.
\end{equation}
The integral can be evaluated exactly when a cut-off $\Lambda$ is used, as usual for Nambu-Jona-Lasinio models. This yields
\begin{equation}
\label{eq:gape}
M_q=m_q+\frac{N_f(N^2-1)Ng^2}{16\pi^2}\sum_{n=0}^\infty\frac{B_nM_q}{2(m_n^2-M_q^2)}
\left[m_n^2\ln\left(1+\frac{\Lambda^2}{m_n^2}\right)-M_q^2\ln\left(1+\frac{\Lambda^2}{M_q^2}\right)\right].
\end{equation}
This equation is amenable to a numerical treatment. It should be solved with the conditions $M_q\ge 0$ and $M_q\ll\Lambda$ that is, the effective mass of the quark should not exceed the ultraviolet cut-off representing, at least, the boundary of the region where asymptotic freedom starts to set in (generally taken at $\Lambda\approx 1\ {\rm GeV}$). We can normalize this equation to the cut-off $\Lambda$ by introducing the new variables $x=m_0/\Lambda$ and $y=M_q/\Lambda$ having taken $m_n=(2n+1)m_0$. The mass $m_0$ can be assumed to be that of the $\sigma$ meson or f(500) that we fix to $m_0=0.417\ {\rm GeV}$. Then,
\begin{equation}
\label{eq:Meff}
y=\frac{m_q}{\Lambda}+\kappa\alpha_s\sum_{n=0}^\infty\frac{B_ny}{(2n+1)^2x^2-y^2}\left[(2n+1)^2x^2\ln\left(1+\frac{1}{(2n+1)^2x^2}\right)-y^2\ln\left(1+\frac{1}{y^2}\right)\right]
\end{equation}
being $\kappa=N_f(N^2-1)N/8\pi$ and $\alpha_s=g^2/4\pi$. We note that the cut-off is completely disappeared except for the ratio $m_q/\Lambda$ that, for the light quarks, is negligible small. The result for eq.(\ref{eq:Meff}), written in implicit form 
\begin{equation}
\label{eq:mu}
\mu(x,y)=y-\frac{m_q}{\Lambda}-\kappa\alpha_s\sum_{n=0}^\infty\frac{B_ny}{(2n+1)^2x^2-y^2}\left[(2n+1)^2x^2\ln\left(1+\frac{1}{(2n+1)^2x^2}\right)-y^2\ln\left(1+\frac{1}{y^2}\right)\right],
\end{equation}
and fixing $x=m_0/\Lambda$ as said above, is given in fig.~\ref{fig:Meff} where the zeros of the function $\mu(x,y)$ are of interest here.
One can see that, when $\alpha_s$ is moving from the asymptotic freedom region, $\alpha_s<1$, toward the infrared fixed point generally taken at $\alpha_s=\pi$, the number of solutions moves from one simple root, very near 0, where the quark retains its small mass, to 3 roots with the quark gaining a dynamical mass and the root near 0 persisting but becoming negative and so, unphysical. Such a scenario is better appreciated in fig.~\ref{fig:Meff2}. 
This implies that, the chiral symmetry, if quarks would be massless, is broken. It is however interesting to note that there is a critical coupling beyond which the effective mass could overcome the cut-off entering into an unphysical region. This is for $\alpha_s\approx 0.5$, given the value of $m_0$ we chose. 

%
Indeed, this particular value of the coupling can be evaluate exactly and can give a confinement condition that applies quite generally to light quarks (u,d,s). This holds when the confinement criteria we stated above applies, in agreement with \cite{Gribov:1998kb, Roberts:1994dr}, that is the on-shell quark mass fails for the propagator. This is given by assuming that that we evaluate eq,~(\ref{eq:mu}) with the condition $y=M_q/\Lambda<1$. So, the value $M_q=\Lambda$ provides a confinement condition. We will get from eq,~(\ref{eq:mu})
\begin{equation}
    \alpha_s=\min_{q=u,d,s}\frac{1-\frac{m_q}{\Lambda}}{\kappa\xi}
\end{equation}
being $\xi$ given by
\begin{equation}
    \xi=\sum_{n=0}^\infty
    \frac{B_n}{(2n+1)^2(m_0/\Lambda)^2-1}\frac{\left[(2n+1)^2(m_0/\Lambda)^2\ln\left(1+\frac{1}{(2n+1)^2(m_0/\Lambda)^2}\right)-\ln 2\right]}{(2n+1)^2(m_0/\Lambda)^2-1}.
\end{equation}
For the values we have chosen, we get $\alpha_s\approx 0.5105$ in agreement with our previous qualitative analysis derived from the breaking of the chiral symmetry. So, in ordinary QCD such a minimum exists and the theory is confining.

\section{Conclusions\label{sec5}}

We have derived the set of Dyson-Schwinger equations for QCD for 1- and 2-point correlation functions. We have seen that they can be cast in a treatable form in the 't~Hooft limit. The main aim of this effort is to present a proof of confinement for the theory. 

%
%
%

Indeed, in the 't~Hooft limit, we were able to show how the transition from the regime of dynamical breaking of symmetry transits to confinement, deriving a general condition for confinement of light quarks that is satisfied by QCD. We hope in future works to extend this analysis.


\begin{thebibliography}{99}



\bibitem{Fodor:2012gf} 
  Z.~Fodor and C.~Hoelbling,
  Rev.\ Mod.\ Phys.\  {\bf 84}, 449 (2012)
  [arXiv:1203.4789 [hep-lat]].

\bibitem{pdg} {\sl Lattice quantum chromodynamics} in M. Tanabashi et al.
(Particle Data Group), Phys. Rev. D 98, 030001 (2018).
	
\bibitem{Bogolubsky:2007ud}
  I.~L.~Bogolubsky, E.~M.~Ilgenfritz, M.~Muller-Preussker, A.~Sternbeck,
  PoS LAT2007, 290 (2007).
  
\bibitem{Cucchieri:2007md}
  A.~Cucchieri, T.~Mendes,
  PoS LAT2007, 297 (2007).
   
\bibitem{Oliveira:2007px}
  O.~Oliveira, P.~J.~Silva, E.~M.~Ilgenfritz, A.~Sternbeck,
  PoS LAT2007, 323 (2007).
	
\bibitem{Lucini:2004my} 
  B.~Lucini, M.~Teper and U.~Wenger,
  JHEP {\bf 0406}, 012 (2004).
	
\bibitem{Chen:2005mg} 
  Y.~Chen {\it et al.},
  Phys.\ Rev.\ D {\bf 73}, 014516 (2006).


\bibitem{Cornwall:1981zr} 
  J.~M.~Cornwall,
  Phys.\ Rev.\ D {\bf 26}, 1453 (1982).
	
\bibitem{Cornwall:2010bk}	
	J.~M.~Cornwall, J.~Papavassiliou, D.~Binosi, 
  ``The Pinch Technique and its Applications to Non-Abelian Gauge Theories'', 	
	(Cambridge University Press, Cambridge, 2010).	

\bibitem{Dudal:2008sp} 
  D.~Dudal, J.~A.~Gracey, S.~P.~Sorella, N.~Vandersickel and H.~Verschelde,
  Phys.\ Rev.\ D {\bf 78}, 065047 (2008)
  [arXiv:0806.4348 [hep-th]].
	  
\bibitem{Frasca:2007uz} 
  M.~Frasca,
  Phys.\ Lett.\ B {\bf 670}, 73 (2008).
	  
\bibitem{Frasca:2009yp}
  M.~Frasca,
  Mod.\ Phys.\ Lett.\  {\bf A24}, 2425-2432 (2009).
	
\bibitem{Frasca:2015yva} 
  M.~Frasca,
  Eur.\ Phys.\ J.\ Plus {\bf 132}, no. 1, 38 (2017)
  Erratum: [Eur.\ Phys.\ J.\ Plus {\bf 132}, no. 5, 242 (2017)]
  [arXiv:1509.05292 [math-ph]].
	
\bibitem{Frasca:2016sky} 
  M.~Frasca,
  Eur.\ Phys.\ J.\ C {\bf 77}, no. 4, 255 (2017)
  [arXiv:1611.08182 [hep-th]].
	
\bibitem{Frasca:2017slg} 
  M.~Frasca,
  Nucl.\ Part.\ Phys.\ Proc.\  {\bf 294-296}, 124 (2018)
  [arXiv:1708.06184 [hep-ph]].
	
\bibitem{Chaichian:2018cyv} 
  M.~Chaichian and M.~Frasca,
  Phys.\ Lett.\ B {\bf 781}, 33 (2018)
  [arXiv:1801.09873 [hep-th]].
	
	%
\bibitem{Eichten:1974et} 
  E.~Eichten and F.~Feinberg,
  Phys.\ Rev.\ D {\bf 10}, 3254 (1974).
	
\bibitem{Baker:1976vz} 
  M.~Baker and C.~k.~Lee,
  Phys.\ Rev.\ D {\bf 15}, 2201 (1977)
  Erratum: [Phys.\ Rev.\ D {\bf 17}, 2182 (1978)].
	
\bibitem{Roberts:1994dr} 
  C.~D.~Roberts and A.~G.~Williams,
  Prog.\ Part.\ Nucl.\ Phys.\  {\bf 33}, 477 (1994)
  doi:10.1016/0146-6410(94)90049-3
  [hep-ph/9403224].

\bibitem{Alkofer:2000wg} 
  R.~Alkofer and L.~von Smekal,
  Phys.\ Rept.\  {\bf 353}, 281 (2001)
  [hep-ph/0007355].

\bibitem{Bender:1999ek} 
  C.~M.~Bender, K.~A.~Milton and V.~M.~Savage,
  Phys.\ Rev.\ D {\bf 62}, 085001 (2000)
  [hep-th/9907045].
	
\bibitem{tHooft:1973alw} 
  G.~'t Hooft,
  Nucl.\ Phys.\ B {\bf 72}, 461 (1974).
	
\bibitem{Frasca:2016rsi} 
  M.~Frasca,
  Eur.\ Phys.\ J.\ C {\bf 78}, no. 9, 790 (2018)
  [arXiv:1602.04654 [hep-ph]].
  
  \bibitem{Smilga:2001ck}	
	A.~V.~Smilga, 
  ``Lectures on quantum chromodynamics'', 	
	(World Scientific, Singapore, 2001).
	
\bibitem{Dyson:1949ha} 
  F.~J.~Dyson,
  Phys.\ Rev.\  {\bf 75}, 1736 (1949).
  doi:10.1103/PhysRev.75.1736
	
\bibitem{Schwinger:1951ex} 
  J.~S.~Schwinger,
  Proc.\ Nat.\ Acad.\ Sci.\  {\bf 37}, 452 (1951).
  doi:10.1073/pnas.37.7.452

\bibitem{Schwinger:1951hq} 
  J.~S.~Schwinger,
  Proc.\ Nat.\ Acad.\ Sci.\  {\bf 37}, 455 (1951).
  doi:10.1073/pnas.37.7.455
  
\bibitem{Frasca:2013tma} 
  M.~Frasca,
  Eur.\ Phys.\ J.\ C {\bf 74}, 2929 (2014)
  doi:10.1140/epjc/s10052-014-2929-9
  [arXiv:1306.6530 [hep-ph]].
  
\bibitem{Gribov:1998kb} 
  V.~Gribov,
  Eur.\ Phys.\ J.\ C {\bf 10}, 71 (1999)
  doi:10.1007/s100529900051
  [hep-ph/9807224].
  
	
\end{thebibliography}
\end{document}